\documentclass[12pt]{article}
\usepackage{epsf}
\usepackage{epsfig}
\usepackage{wrapfig}
\usepackage{amsmath}
\usepackage{amssymb}
\usepackage{eufrak}
\usepackage{graphicx}
\newcommand{\Tr}{{\rm Tr}}
\begin{document}

\begin{flushright}
\end{flushright}
\vskip 15mm

\begin{center}
\section*{New Kinds of Phase Transitions: Transformations in disordered
Substances}

\vskip 5mm
V.N. Ryzhov$^{1}$ and E.E. Tareyeva$^{1}$
\vskip 5mm

{\small
(1) {\it Institute for High Pressure Physics, Russian Academy of
Sciences, Troitsk 142190, Moscow region, Russia}
}
\end{center}
\vskip 5mm

\begin{abstract}
The transitions in disordered substances are discussed briefly:
liquid--liquid phase transitions, liquid--glass transition and the
transformations of one amorphous form to another amorphous form of the
same substances. A description of these transitions in terms of
many--particle conditional distribution functions is proposed.
The concept of a hidden long range order is proposed, which is
connected with the broken symmetry of higher order distribution
functions. The appearance of frustration in simple supercooled
Lennard--Jones liquid is demonstrated.

\end{abstract}
\vskip 8mm

It is well known for a long time that there exist sharp phase transitions
between different aggregate states and sharp polymorphic phase transitions
between different crystalline phases of the same substances. During last
two decades a lot of experimental data was obtained on complicated phase
diagrams of liquids and amorphous solids, too. Some of these results were
presented on the first international conference that took place here in
Russia in 2001 \cite{book} and that can be considered as a formal claim of
a new direction of physical investigations -- transformations in
disordered substances: liquid--liquid transitions and transformations of
one amorphous form to another amorphous form of the same substances. The
liquid--glass transition, although has longer history, has to be considered
in the same context.
It is now firmly established by different experimental techniques
that sharp liquid--liquid transitions under
pressure, formally similar to first--order phase transitions, exist
as well as reversible transformations between amorphous states
involving changes in local order structures and density. Usually a crystal
melts with a conservation of the short--range order (SRO) structure type
or into denser liquid with SRO structure similar to that of high pressure
crystalline phase.

It should be emphasized that
the transitions in liquids are true phase transitions mainly determined by
thermodynamic relationships, whereas the transitions in amorphous solids
take place far away from equilibrium and are governed by the corresponding
kinetics.

A useful microscopic theory of these transitions is not developed
yet and only empirical models and computer simulations have been used in
practice to date.
For example, interesting results
were obtained by Stanley through molecular dynamics study basing on the
taking into account the hydrogen bonding in supercooled water
\cite{stanley1}. It is by the demonstration of a simple analytic way
of obtaining Stanley results   that we begin the
presentation of our own results on this subject (see, e.g.,
\cite{book,ryz79,ryz1,pn,clust} and references therein).

From the intuitive point of view liquid-liquid phase transition between
low density and high density phases may be related to the competition
between expanded and compact structures. This suggests that the potential
should have two equilibrium positions. The most obvious form of such
potential is:
\begin{equation}
\Phi(r)=\left\{
\begin{array}{ll}
\infty, & r\leq \sigma\\
0, & \sigma<r\leq a\\
-\varepsilon_1, & a<r\leq b\\
0, & b<r\leq c\\
-\varepsilon_2, & c<r\leq d.
\end{array}\right..
\label{1}
\end{equation}
This two--well potential may
be considered as a model for the water potential
\cite{stanley1}.

To investigate the possibility
of the existence
of the second critical point in this case
we developed the mean-field
(van der Waals--like) theory.
Using the well-known Bogoliubov inequality for the free energy we can write
$F\leq F_{HS}+<U-U_{HS}>_{HS}$. Here $F_{HS}$ is the free energy
of the system of hard spheres with diameter $\sigma$,
and we consider the attractive part as a
perturbation. Here $U=\frac{1}{2}\sum_{i\neq j}^N\Phi(r_{ij})$ and
$U_{HS}=\frac{1}{2}\sum_{i\neq j}^N\Phi_{HS}(r_{ij})$. The average over
the hard sphere potential has the form
\begin{equation}
<U-U_{HS}>_{HS}=2\pi\rho N\int_0^{\infty}\Phi_{atr}(r)g_{HS}(r)r^2dr,
\label{2}
\end{equation}
where $\Phi_{atr}(r)=\Phi(r)-\Phi_{HS}(r)$,
$g_{HS}(r)$ is the radial distribution function of
the hard sphere system which we take in the Percus-Yevick approximation
\cite{henderson} and for
$F_{HS}$ we use
the approximate Carnahan-Starling equation \cite{barker2}
\begin{equation}
\frac{F_{HS}}{k_BTN}=3\ln\lambda-1+\ln\rho+\frac{4\eta-3\eta^2}{(1-\eta)^2}.
\label{3}
\end{equation}
Here $\lambda=h/(2\pi mk_BT)^{1/2}$.

\begin{figure}
\begin{center}
\includegraphics[width=8cm]{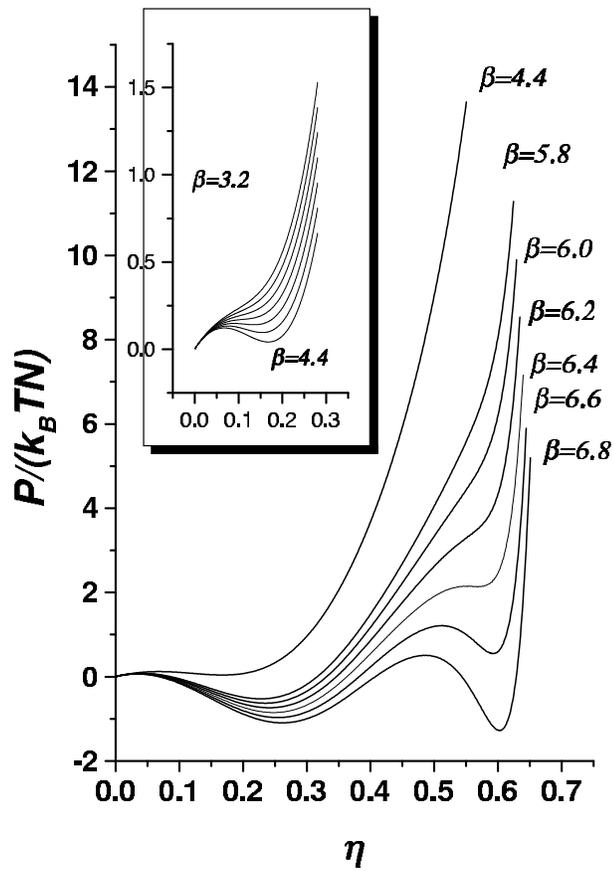}
\caption{The isotherms for the double-well potential (1) for different values
of $\beta=\varepsilon_1/(k_B T)$.}
\end{center}
\end{figure}

The equation of state is given by
$P=\rho^2\partial(F/N)/\partial\rho$. In Fig. 1 the two families
of isotherms are shown for the temperatures close to two critical
points $\beta_1=\varepsilon_1/k_BT_{c1}$ and
$\beta_2=\varepsilon_1/k_BT_{c2}$. Fig. 1 shows that at low
temperatures ($\beta=\varepsilon_1/k_BT>\beta_1$) there are two
van der Waals--like loops in the equation of state which correspond
to two fluid-fluid transitions. In the temperature region
$\beta_2<\beta<\beta_1$ there is only one loop which corresponds
to the well known gas-liquid transition, $\beta_2$ being the
gas-liquid critical point temperature and $\beta_1$ -- the
liquid-liquid critical temperature.

In this example, as at the
ordinary critical point, no symmetry of the correlation functions
is broken at the transition. The order parameter is
the difference of densities of high and low density phases
$\Delta\rho=\rho_{l1}-\rho_{l2}$. However, it is interesting to describe
the change of the local structure and the cluster symmetry at the
transition.

To describe different kinds of space symmetry breaking we use
the formalism of classical many particle conditional
distribution functions
\[ F_{s+1}({\bf r}_1|{\bf r}_1^0 ... {\bf r}_s^0)=
\frac{F_{s+1}({\bf r}_1, {\bf r}_1^0,...,{\bf r}_s^0)}
{F_s({\bf r}_1^0,...,{\bf r}_s^0)}. \]
Here $F_s({\bf r}_1,...,{\bf r}_s)$ is the usual $s$--particle distribution
function \cite{NNB2}. The functions $F_{s+1}({\bf r}_1|{\bf r}_1^0 ...
{\bf r}_s^0)$ satisfy the equation
\begin{eqnarray}
\frac{\rho F_{s+1}({\bf r}_1|{\bf r}_1^0 ... {\bf r}_s^0)}{z}& = &\exp
\left\{ -\beta \sum_{k=1}^s \Phi({\bf r}_1-{\bf r}_k^0) +\sum_{k \geq 1}
\frac{\rho^k}{k!} \int \, S_{k+1}({\bf r}_1,...,{\bf r}_{k+1}) \right.
\nonumber \\
& &\left.\times F_{s+1}({\bf r}_|{\bf r}_1^0 ... {\bf r}_s^0)...
F_{s+1}({\bf r}_{k+1}|{\bf r}_1^0 ... {\bf r}_s^0)
d{\bf r}_2... d{\bf r}_{k+1} \right\} \label{main}.
\end{eqnarray}
Here
$z $ is activity, $\rho$ is the mean number density, $S_{k+1}({\bf
r}_1,...,{\bf r}_{k+1})$ is the irreducible cluster sum of Mayer functions
connecting (at least doubly) $k+1$ particles,
$\beta=1/k_B T$.

The simplest case is the symmetry breaking of the one--particle
function. In the solid phase the local density, proportional to the
one-particle distribution function, has the symmetry of a crystal lattice
and can be expanded in a Fourier series in reciprocal lattice vectors ${\bf
G}$:
\begin{equation}
\rho({\bf r})=\sum_{\bf G}\rho_{\bf G}e^{i{\bf Gr}},
\label{4}
\end{equation}
where the Fourier coefficients $\rho_{\bf G}$ are the order
parameters for the transition.

The Taylor expansion of the corresponding free energy functional around the
liquid can be written in the following form:
\begin{equation} \beta \Delta F = \int
d{\bf r} \varrho ({\bf r}) \ln \frac {\varrho ({\bf r})}
{\varrho _0} - \sum_{k \geq 2} {1 \over k!} \int c^{(n)} ({\bf
r}_1,...,{\bf r}_k) \Delta \varrho ({\bf r}_1)...\Delta\varrho
({\bf r}_k) d{\bf r}_1 ... d{\bf r}_k , \label{exfree}
\end{equation}
where
$$ \Delta \varrho ({\bf r}) = \varrho ({\bf r}) - \varrho_l $$
is the local density difference between solid and liquid phases.
This is the base of the density functional theory of freezing (DFT)
\cite{ryz79}. In the frame of this approach tens
of melting curves were calculated (see, e.g., the reviews \cite{rev}).
The full system of equations to be solved in  DFT contains the nonlinear
integral equation for the function $\rho ({\bf r})$, obtained as the
extremum condition for the free energy and the equilibrium conditions for
the chemical potential and  the pressure written in terms of the same
functions as in (\ref{exfree}). To proceed constructively in the frame of
 DFT we have to choose an actual form of the free energy functional -- a
kind of closure or truncating -- and we must make an ansatz for the
average density of the crystal. The importance of such an ansatz follows
from the fact that we are dealing with a theory which is equivalent to
Gibbs distribution and one has to break symmetry following the Bogoliubov
concept of quasiaverages \cite{bogol1}.

Now let us consider a state of matter which is characterized
by the uniform local density, but the broken symmetry
of the two--particle distribution function. Such type of
order is called the bond orientations order (BOO), where ``bond'' is the
vector joining a particle with its nearest neighbor. This kind of order is
well known in theories of two-dimensional melting (hexatic phase)
\cite{halpnel79,ryz1}.
Near the transition to anisotropic liquid state we have:
\begin{equation}
F_2({\bf r}_1|{\bf r}_1^0)=g(|{\bf r}_1-{\bf r}_1^0|)
(1+f({\bf r}_1-{\bf r}_1^0)), \label{F2bo}
\end{equation}
where  $f({\bf r}_1-{\bf r}_1^0)$ has the symmetry of the local environment
of the particle at  ${\bf r}_1^0$ and may be written in the form
$f({\bf r}_1-{\bf r}_1^0)=f(|{\bf r}_1-{\bf r}_1^0|,\Omega),$
$\Omega$ determines the direction of the vector
${\bf r}_1-{\bf r}_1^0$. In the case of three dimensions function
$f(r,\Omega)$ may be expanded in a series in spherical harmonics:
\begin{equation}
f(r,\Omega)=\sum_{l=0}^{\infty}\sum_{m=-l}^{l}f_{lm}(r)Y_{lm}(\Omega).
\label{3d}
\end{equation}

The microscopic equations for the order parameters $f_{lm}(r)$ can be
obtained from the main equation (\ref{main}). The linearized equation
determines the instability of isotropic liquid against the formation of
the state with BOO and has the form \cite{ryz1}:
\begin{equation}
f_{lm}(r)-\frac{4\pi}{2l+1}\int\,\Gamma_l(r,r')g(r')f_{lm}(r')r'^2\,dr'=0.
\label{inst}
\end{equation}
Here $\Gamma_l(r,r')$ correspond to the isotropic liquid when
\begin{eqnarray}
\Gamma({\bf r}_1,{\bf r}_1^0,{\bf r}_2)&=&\sum_{k\geq 1}\frac{\rho^k}
{(k-1)!}\int\,S_{k+1}({\bf r}_1...{\bf r}_{k+1})\times \nonumber\\
&\times&g(|{\bf r}_3-{\bf r}_1^0|)\cdots g(|{\bf r}_{k+1}-{\bf r}_1^0|)\,
d{\bf r}_3 \cdots d{\bf r}_{k+1}. \label{Gamma}
\end{eqnarray}
reduces to
\begin{equation}
\Gamma({\bf r}_1,{\bf r}_1^0,{\bf r}_2)=
\Gamma(r,r',\theta), \label{K}
\end{equation}
\begin{equation}
\Gamma(r,r',\theta)=\sum_{l=0}^{\infty}\frac{4\pi}{2l+1}\Gamma_l(r,r')
\sum_{l=-m}^{l}Y_{lm}(\Omega_1)Y_{lm}^*(\Omega_2), \label{Gexp}
\end{equation}
The angles $\Omega_1$ and $\Omega_2$ determine the directions of the vectors
${\bf r}$ and ${\bf r'}$ and
$r=|{\bf r}_1-{\bf r}_1^0|,  r'=|{\bf r}_2-{\bf r}_1^0|,  \theta$
is the angle between vectors ${\bf r}$ and ${\bf r'}$.
It should be notice that
the correlation length of the orientational fluctuations
$\xi_{l,m}\rightarrow\infty$ when approaching the instability line
given by Eq. (\ref{inst}).

To describe liquid--liquid and liquid--glass transitions we must consider
isotropic case with rotationally invariant two--particle distribution
function. A possible description of these cases can be given
in terms of broken symmetry of higher order distribution functions.
At high temperature the nearest neighbors of a molecule can take
different relative positions and there is no SRO.
At lower temperature SRO appears which can be of
different kinds at different densities.
The rotation and
the translation of the clusters of preferred symmetry give rise to
the fact that one-particle and two-particle distribution
functions remain isotropic. If a kind of BOO appears the clusters are
oriented in similar way and the
two-particle distribution function becomes to be anisotropic
(as in 2D hexatic phase). However, we can imagine another
situation -- freezing of the symmetry axes of the clusters in
different position. The isotropic phase can be considered as
analogous to the paramagnetic phase (of cluster symmetry axes),
the BOO phase -- to the ferromagnetic phase, and the mentioned
freezed phase -- to a spin glass phase.

Let us consider for simplicity a 2D system. In the vicinity of the
transition one can write (in the superposition approximation for
the liquid)
 \begin{equation}
F_3({\bf r}_1| {\bf r}_1^0, {\bf r}_2^0) =
g(|{\bf r}_1-{\bf r}_1^0|) g(|{\bf r}_1-{\bf r}_2^0|)(1+f_3({\bf
r}_1| {\bf r}_1^0, {\bf r}_2^0) \label{gla}
\end{equation}
In 2D case
$f_3({\bf r}_1| {\bf r}_1^0, {\bf r}_2^0)$ depends in fact
on two distances and two angles
 \begin{equation}
f_3({\bf r}_1| {\bf r}_1^0, {\bf r}_2^0) =
f_3(R_0, \phi _0;R_1, \Theta _1 ),
\label{gla1}
\end{equation}
where
$ {\bf R}_0 = {\bf r}_2^0 - {\bf r}_1^0$,
$ {\bf R}_1 = {\bf r}_1 - {\bf r}_1^0$,
$ {\bf R}_2 = {\bf r}_2 - {\bf r}_1^0$ and $\phi _0$ is the
angle of the vector ${\bf R}_0$ with the $z$ axis,
$ \Theta _1$ -- the angle between ${\bf R}_1$ and ${\bf R}_0$
and $ \Theta _2$ -- the angle between ${\bf R}_2$ and ${\bf
R}_0$.

The linearization of ~(\ref{main})  for $s=2$ gives:
\begin{equation}
f_3(R_0, \phi _0;R_1, \Theta _1)=\int \,
\Gamma'(R_0, \phi _0;{\bf r}_2; R_1, \Theta _1)
f_3(R_0, \phi _0;R_2, \Theta _2)
g(|{\bf R}_2-{\bf R}_0|) g(R_2) d{\bf r}_2,
\label{gla2} \end{equation}
where
\begin{eqnarray}
\Gamma'(R_0, \phi _0;{\bf r}_2; R_1, \Theta _1)&=&
 \sum_{k \geq 1}
 \frac{\rho^{k}}{(k-1)!}\, \int\, S_{k+1}({\bf r}_1,...,{\bf
 r}_{k+1})g(|{\bf r}_3-{\bf r}_1^0|)\, \nonumber\\ &\times&
g(|{\bf r}_3-{\bf r}_2^0|)...
g(|{\bf r}_{k+1}-{\bf r}_1^0|)g(|{\bf r}_{k+1}-{\bf r}_2^0|) \,
d{\bf r}_3...d{\bf r}_{k+1}.  \label{gla3}
\end{eqnarray}
There are two kinds of angles entering the equations and two kinds
of order parameters, consequently. One angle ($\phi _0$) fixes the
position of one pair of particles of the cluster, and the other
($\Theta _i$) -- the position of the third particle in the
coordinate frame defined by $\phi _0$. The order parameter
connected with $\Theta _i$ is the generalization of intracluster
hexatic parameter for the case of different coordinate frames. The
order parameter connected with $\phi _0$ is an analogue of
magnetic moment and in glass--like phase one can consider an
Edwards-Anderson parameter $<\cos \phi  _0 (t) \cos \phi _0(0)>$.
In such a way we come to the concept of a ``conditional'' or ``hidden''
long range order: if we consider two pairs of particles at infinite
distance from one another then there exists a preferable possibility for
the relative position of the third particle near each pair. The directions
of the bonds in the pairs of particles themselves are subjects to
spin--glass--like order.  In 3D case the rotation of clusters is given by
rotation matrices $D_{lm}^{l'm'}(\vec \omega _{0i})$ so that we obtain a
kind of orientational multipole glass for the clusters. If the
intracluster ordering is established then we can consider the system of
clusters. The orientational state of this system is defined by the
intercluster interaction for different values of temperature an pressure.

Now let us consider this later situation when the intracluster symmetry is
fixed and let us try to estimate the intercluster
orientational interaction. If the intercluster interaction had the same
sign for all cluster sizes (or all clusters had the same size) one would
get the state with simple BOO. However, because of the difference in
cluster sizes the orientational interaction for some harmonics may change
sign as a function of the cluster size (see Fig. 2). In this case the low
temperature state should be amorphous for some harmonics.
So the difference of the
orientational interaction of the clusters for different cluster sizes may
be considered as the reason of some kind of frustration in simple liquids.
It should be emphasized that the form of the corresponding component of
the orientational interaction is the intrinsic statistical property of the
liquid and does not depend on the timescale of the fluctuations in size
and symmetry of the cluster. There is no real quenched disorder in the
system but only an analog of it which may be treated in a formally same
way as the quenched disorder in spin glasses. To analyze qualitatively the
orientational freezing in the system we introduce simple lattice model
which takes into account the interaction only between clusters with
definite symmetry. The model gives the possibility to conclude what
harmonics freeze first and what local symmetry prevails immediately below
the transition. Let us now describe our results in more detail.

Our starting point is the expression for the free energy of the system as a functional of a
pair distribution function $g_2({\bf r}_i,{\bf r}_0)$ which has the form
\cite{ryz1}:
\begin{eqnarray}
&&F/k_BT=\int d{\bf r}d{\bf r}_0 \rho g_2({\bf r},{\bf r}_0)\left[\ln\left(\lambda^3
\rho g_2({\bf r},{\bf r}_0)\right)-1\right]- \nonumber\\
&&-\sum_n \frac{\rho^{n+1}}{(n+1)!}\int S_{n+1}({\bf
r}_1...{\bf r}_{n+1})
g_2({\bf r}_1,{\bf r}_0)\cdots g_2({\bf r}_{n+1},{\bf r}_0) \times \nonumber\\
&&\times d{\bf r}_1\cdots d{\bf r}_{n+1}d{\bf r}_0
-\int \Phi({\bf r}-{\bf r}_0)\rho g_2({\bf r},{\bf r}_0) d{\bf r}d{\bf r}_0, \label{1}
\end{eqnarray}
where the term  with logarithm corresponds to the entropy
and the other terms
--- to the interaction energy.
Here $\Phi({\bf r}-{\bf r}_0)$ - interparticle potential (for
Lennard-Jones potential, $\Phi_0(r)=4\varepsilon((\sigma/r)^{12}-(\sigma/r)^6)$),
$\lambda=h/(2\pi mk_BT)^{1/2}$.

We can
estimate the change of the energy due to (\ref{F2bo}).
Omitting the entropy term in Eq.\ref{1}, we have
up to the second order in
$\delta g({\bf r},{\bf r}_0)$.
\begin{eqnarray}
\Delta F/k_bT&=&-\frac{1}{2}\int \Gamma({\bf r}_1,{\bf r}_0,{\bf
r}_2)
\delta g({\bf r}_1,{\bf r}_0)\delta g({\bf r}_2,{\bf r}_0)
d{\bf r}_1 d{\bf r}_2.\label{5}
\end{eqnarray}
Using the approximation for the
radial distribution function $g(r)=\rho^{-1}(n_s/4\pi r_s^2)\delta(r-r_s)$
and the Eq.(\ref{3d}) we obtain:
\begin{eqnarray}
&&\Delta
F(r_s)/k_BT=-\frac{1}{2}\rho^{-2}\left(\frac{n_s}{4\pi}\right)^2\sum_{l=0}^{\infty}\frac{4\pi}{2l+1}
\Gamma_l(r_s,r_s)\sum_{m=-l}^{l}%\nonumber\\
%&&\times\int Y_{lm}(\Omega_1)Y_{lm}^*(\Omega_2)
\int Y_{lm}(\Omega_1)\times\nonumber\\
&&\times Y_{lm}^*(\Omega_2)
f(r_s,\Omega_1)f(r_s,\Omega_2) d\Omega_1d\Omega_2=%\nonumber\\
-\frac{1}{2}\sum_{l=0}^{\infty} J_l(r_s)\sum_{m=-l}^l |f_{lm}|^2.
\label{7}
\end{eqnarray}
Here $J_l(r_s)=\rho^{-2}\frac{4\pi}{2l+1}(\frac{n_s}{4\pi})^2\Gamma_l(r_s,r_s)$, $n_s$ is the
number of nearest neighbors of a particle  and $r_s$ is the size of the
cluster, which is of the order of the first coordination shell size.

The function $\Delta F(r_s)$ may be interpreted as the mean-field
orientational interaction energy of the system of clusters having the size
$r_s$. To get the full energy of the system one should integrate (\ref{7})
over the probability of finding the cluster with the size $r_s$ which is
given by the function $r_s^2g(r_s)$ in the vicinity of the first maximum.

Using the approximations of \cite{ryz1} for
$\Gamma({\bf r}_1,{\bf r}_1^0,{\bf r}_2)$
we obtain the estimation for $J_l(r_s)$ as a function of $r_s$.
Fig.2 represents $J_l(r_s)$ for $l=4$ and $6$ along with $r_s^2g(r_s)$ in
the vicinity of its first peak. It is seen that $J_l(r_s)$ changes sign.
This result enables us to suppose that there is a kind of frustration
(which is analogous to that in spin glasses) appearing as a result of
variations in the sizes of clusters according to $g(r)$ and that it is
possible
to study the transition in the system of interacting clusters on the base
of simple model lattice Hamiltonian:
\begin{equation}
H=-\frac{1}{2}\sum_{<i\neq
j>}\sum_{l=0}^{\infty}J^l_{ij}\sum_{m=-l}^l
U_{lm}(\Omega_i)U_{lm}^*(\Omega_j). \label{8}
\end{equation}
The functions $U_{lm}(\Omega_i)$ are the lattice harmonics for the point groups corresponding
to the cluster symmetry. This Hamiltonian describes correctly
BOO of clusters. In this case the energy
calculated from Eq. (\ref{8}) in the mean-field approximation (taking into account that
$\langle U_{lm}(\Omega_i)\rangle=f_{lm}$) coincides with the intercluster energy (\ref{7})
under appropriate choice of $J^l_{ij}$. We will use the Hamiltonian (\ref{8}) to study the
system of interacting clusters with various sizes.

\begin{figure}
\begin{center}
\includegraphics[width=8cm]{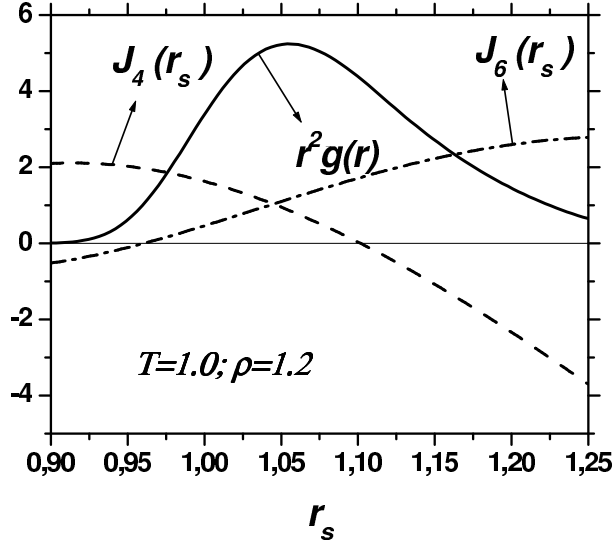}
\caption{$J_l(r_s)$ for $l=4$ and $6$ along with $r_s^2g(r_s)$ in the vicinity of
the first peak as functions of $r_s$ for dimensionless temperature $k_BT/\varepsilon=1.0$ and
density $\rho\sigma^3=1.2$. }
\end{center}
\end{figure}

To simplify the problem we neglect in Hamiltonian (\ref{8}) all the terms except ones
corresponding to the unit representation of the point group. Furthermore, we consider only the
cases $l=4$ and $l=6$ which represent the cases of cubic and icosahedral symmetries. This
Ising-like model may be called a ``minimal'' model:
\begin{equation}
\label{one} H=-\frac{1}{2}\sum_{i\neq j}J_{ij}
\hat{U_i}\hat{U_j}.
\end{equation}
Functions $\hat U\equiv U(\varphi,\theta)$ are the combinations of spherical harmonics. We
will consider separately symmetries of ``simple'' cube ($l=4, m=0,\pm 4$), cube ($l=6, m=0,\pm
4$) and icosahedron ($l=6, m=0,\pm5$) correspondingly \cite{hay1,Bredli}. For example, for
$l=4$ one has:
\begin{eqnarray}
\hat U\equiv
U(\varphi,\theta)=\sqrt{\frac{7}{12}}\left\{Y_{40}(\varphi,\theta)+
\sqrt{\frac{5}{14}}\left(Y_{44}(\varphi,\theta)+
Y_{44}(-\varphi,\theta)\right)\right\}
\end{eqnarray}

The interactions $J_{ij}$ are chosen in such a way that the MF approximation gives exact
solution (infinite-range interactions). It is easily seen that in the
minimal model (\ref{one}) without disorder in the framework of the MF
approximation there is a first order phase transition to the state with
BOO (compare to \cite{hay1,Nelson-book}). From Fig.2 it is clear that, as
the first qualitative step, $J^l_{ij}$ may be chosen as random
interactions with Gaussian probability distribution
\begin{equation}
\label{two}
P(J_{ij})=\frac{1}{\sqrt{2\pi
J}}\exp\left[-\frac{(J_{ij}-J_0)^{2}}{ 2J^{2}}\right]
\end{equation}
where $ J=\tilde{J}/\sqrt{N}$ , $J_{0}=\tilde{J_0}/N$ can be related to the
microscopic parameters. We approximate $r^2 g(r)$ by a gaussian
exponential near the position of the first maximum $r_0$. So $r^2g(r)\sim
\exp[-(r-r_0)^2/2\sigma]$. The approximation for the functions $\Gamma_l$
is then linear: $\Gamma_l\approx \alpha+\beta(r-r_0)$. That is:
$J_0=\alpha, J=\beta\sqrt\sigma$.

The free energy of the system can be obtained using replica approach
 (see, e.g., \cite{SK}). In the
 replica-symmetric (RS) approximation we have \cite{4avtora}:
\begin{eqnarray}
\label{four}
F=-NkT\biggl\{-\left(\frac{\tilde{J_0}}{kT}\right)\frac{m^2}{2}+
t^2\frac{q^2}{4}-t^2\frac{p^2}{4}+\nonumber\\
\int_{-\infty}^{\infty}\frac{dz}{\sqrt{2\pi}}\exp\left(-\frac{z^2}{2}\right)\ln
\Tr\left[\exp\left(\hat\theta\right)\right]\biggr\},
\end{eqnarray}
where the trace in this case is defined as follows: $\Tr(\ldots)\equiv \int_0^{2\pi}d\varphi
\int_0^\pi d\cos(\theta)(\ldots)$.  Here $t=\widetilde{J}/k_BT$ and
$$\hat{\theta}=\left[zt\sqrt{q}+m\left(\frac{\tilde{J_0}}{kT}\right
)\right]\hat{U}+t^2\frac{p-q}{2}\hat{U}^2.$$

The order parameters are: $ m $ is the regular order parameter (an analog of magnetic moment
in spin glasses), $ q$ is the glass order parameter and $p$ is an auxiliary order parameter.
The extremum conditions for the free energy ~(\ref{four}) give the following equations for
these order parameters:
\begin{equation}
m=\overline{\langle \hat U\rangle},\qquad
p=\overline{\langle \hat U^2\rangle},\qquad
q=\overline{\langle \hat U\rangle^2} \label{prs},
\end{equation}
where $\langle\ldots\rangle=\Tr( \ldots e^{\hat\theta})/\Tr
e^{\hat\theta}$ and $\overline{(\ldots)}=\int_{-\infty}^{\infty}
\frac{dz}{\sqrt{2\pi}}e^{-z^2/2}[\ldots]$.
We find from these equations the temperature dependence of the order
parameters. The RS solution is stable unless the replicon mode
energy $\lambda_{\rm repl}$ is nonzero \cite{A-T,4avtora}. For our model
we have
\begin{equation}
\lambda_{\rm repl}=1-t^2\overline{\langle\langle \hat U^2
\rangle\rangle^2},
\end{equation}
where $\langle\langle\ldots\rangle\rangle$ denotes the irreducible correlator. We find the
temperature $T_{_{A-T}}$ that corresponds to $\lambda_{\rm repl}=0$. To obtain the actual
glass transition temperature one has to study the dynamics of the system. In this paper we
limit ourselves by the static approach. As is usually believed \cite{cugl,KT} and correctly
shown in \cite{Kurchan} the dynamical $T_g$ can be obtained in the frame of the static
approach as the temperature $T_m$ of the marginal instability of the
one-step RS breaking solution. We have calculated $T_m$ and found that
within the accuracy of calculations $T_m$ and $T_{_{A-T}}$ coincide. We
expect that as in spin glasses below $T_{_{A-T}}$ the liquid  dynamics is
characterized by long relaxation times and other phenomena characteristic
to glass transitions. So there is the glass transition in the simple cube
case with $T_{_{A-T}}\approx 0.39$; in the other cases, icosahedron and
cube, there is no glass transition but just a first order transition to
BOO state at temperatures about $0.45, 0.42$
correspondingly at $\rho\sigma^3=0.973$. The last two temperatures of
BOO transitions are in agreement with the results of
molecular dynamics simulations of Ref.\cite{st83}. It should be noted that
all these temperatures are well below the melting temperature $T=0.703$ at
this density \cite{st83,Nelson-book}.

The work was supported in part by the Russian Foundation for Basic Research
(Grant No 02-02-16622 (VNR), Grant No 02-02-16621 (EET) and RFBR-NWO Grant
No 04-01-89005 (047.016.001.).


\begin{thebibliography}{99}
\itemsep=-2pt
\bibitem{book} V.V. Brazhkin, S.V. Buldyrev, V.N. Ryzhov,
and H. E. Stanley [eds], {\it New Kinds of Phase Transitions: Transformations
in Disordered Substances} [Proc. NATO Advanced Research Workshop, Volga River]
(Kluwer, Dordrecht), 2002.

\bibitem{stanley1}
O. Mishima and H. E. Stanley, Nature {\bf 396}, 329 (1998).

\bibitem{ryz79} V. N. Ryzhov and E. E. Tareyeva, Phys. Lett. A
{\bf 75}, 88 (1979).

\bibitem{ryz1}
V. N. Ryzhov and E. E. Tareyeva, Theor. Math. Phys. {\bf 73}, 463 (1987);
J. Phys. C: Solid State Phys. {\bf 21},819 (1988).
Phys. Rev. B {\bf 51},8789 (1995); JETP {\bf 81}, 1115 (1995).

\bibitem{pn}
E.E. Tareyeva and V.N. Ryzhov, Particles\&Nuclei {\bf 31}, part 7B,
184 (2000).

\bibitem{clust} N.M.Chtchelkatchev, V.N.Ryzhov, T.I.Schelkacheva,
E.E.Tareyeva, Phys.Lett. A {\bf 329}, 244 (2004).

\bibitem{henderson}
W. R. Smith and W. Henderson, Mol. Phys. {\bf 19}, 411 (1970).

\bibitem{barker2}
J.A. Barker and D. Henderson, Rev. Mod. Phys. {\bf 48}, 587 (1976).



\bibitem {NNB2} N.N.Bogoliubov, {\it Problems of dynamical
theory in statistical physics} (Moscou, Gostehisdat, 1946).%47

\bibitem{dft}
T.V.Ramakrishnan, M.Youssouff, Phys. Rev.
{\bf B 19}, 2775 (1979);
A. D. J. Haymet and D. W. Oxtoby, J. Chem. Phys. {\bf 74},
2559 (1981).

\bibitem{rev}
 Y. Singh, Phys. Rep. {\bf 207}, 351 (1991);
 M. Baus, J. Phys.: Condens. Matter {\bf 1}, 3131
(1989); H.L\"owen, Phys.Rep. {\bf 237}, 249 (1994).

\bibitem{bogol1} N.N.Bogoliubov, JINR Preprint R-1451, Dubna, 1963;
Phys.Abh.S.U., {\bf 6}, 1, 113, 229 (1962).

\bibitem{halpnel79}
D.R. Nelson and B.I. Halperin, Phys. Rev. B {\bf 19}, 2457 (1979).

\bibitem{hay1} A.D.J. Haymet, Phys. Rev. B {\bf 27}, 1725 (1983).

\bibitem{Nelson-book} D.R. Nelson, ``Deffects and geometry in condensed
matter physics'', Cambridge University Press, 2002.

\bibitem{Bredli} C.J. Bradley, A.P. Cracnell,
\textit{The mathematical theory of symmetry in solids}, (Clarendon, Oxford), 1972.

\bibitem{4avtora}
N.V. Gribova, V.N. Ryzhov, T.I. Schelkacheva, E.E.
Tareyeva. Phys. Letters A, {\bf 315}, 467 (2003).

\bibitem{A-T} J.R.L. Almeida and D.J. Thouless,
J.Phys. A: Math. Gen. {\bf 11}, 983 (1978).

\bibitem{cugl} L.F. Cugliandolo, {\it Dynamics of glassy systems},
arXiv: cond-mat/0210312.

\bibitem{KT} T.R. Kirkpatrick and D. Thirumalai, Phys. Rev. Lett.
{\bf 58}, 2091 (1987).

\bibitem{Kurchan} A. Baldassari, L.F. Cugliandolo, J. Kurchan, and G. Parisi, J. Phys. A:
Math. Gen. {\bf 28}, 1831 (1995).

\bibitem{st83}
P.J. Steinhardt, D.R. Nelson, and M. Ronchetti, Phys. Rev. B {\bf 28},
784 (1983).

\bibitem{SK} D. Sherrington and S. Kirkpatrick, Phys. Rev. Lett. {\bf 32},
1972 (1975); S. Kirkpatrick, D. Sherrington, Phys. Rev.  {\bf
B17}, 4384 (1978).
\end{thebibliography}
\end{document}